\documentclass[a4paper,11pt]{article}
\usepackage{pos}
\makeatletter\let\ltx@label\label\makeatother
\usepackage{lineno}
\title{Jet-like di-hadron correlations in pO, OO, and Ne--Ne collisions}

\author*[\ref{aaa}]{Suraj Prasad}
\affiliation{HUN-REN Wigner Research Centre for Physics, 29-33 Konkoly-Thege Mikl\'os Road, 1121 Budapest, Hungary}
\emailAdd{suraj.prasad@cern.ch}
\note{for the ALICE Collaboration\label{aaa}}
\abstract{
Heavy-ion collisions are primarily studied to investigate the properties of hot and dense nuclear matter, known as the quark--gluon plasma (QGP). However, recent observations of hydrodynamic flow-like behaviour in pp and p--Pb collisions at the LHC suggest the possible formation of QGP droplets even in small collision systems. To bridge the multiplicity gap between small and large systems and to examine the applicability of hydrodynamic descriptions, light-ion collisions were performed for the first time at the LHC in 2025.
This contribution presents measurements of the near-side peak width of di-hadron correlations in pO, OO, and Ne--Ne collisions with ALICE, used to search for medium effects. The width of the near-side correlation peak is studied as a function of the average charged-particle multiplicity at midrapidity. At low transverse momentum ($p_{\rm T}$), the near-side peak is broader in OO and Ne--Ne collisions than in pp and pO collisions, and in central collisions its width approaches that measured in Pb--Pb collisions. At higher $p_{\rm T}$ the peak narrows modestly with increasing multiplicity. The results are compared with PYTHIA, AMPT, and JETSCAPE calculations, none of which reproduces the measured widths over the full kinematic range.
}

\FullConference{14th Edition of the Large Hadron Collider Physics (LHCP2026)\\
18-22 May 2026\\
Paris, France\\}

\begin{document}
\maketitle

\section{Introduction}
\label{sec:intro}
Heavy-ion collisions at RHIC and the LHC produce a hot and dense state of deconfined partons known as the quark--gluon plasma (QGP). The formation of QGP is usually inferred from collective flow and the energy loss of energetic partons caused by their interactions with the QGP medium~\cite{ALICE:JourneyQCD,Busza:2018rrf,Connors:2017ptx}. Over the past decade, small collision systems have shown several signatures similar to those observed in heavy-ion collisions, raising questions about the extent to which QGP-like phenomena can emerge in such systems. Ridge-like long-range correlations and non-vanishing flow coefficients have also been measured in high-multiplicity pp and p--Pb collisions~\cite{Nagle:2018nvi,GrosseOetringhaus:2024ces,ALICE:PartonicFlow}. Parton energy loss, on the other hand, has not been observed unambiguously in these systems~\cite{ALICE:2017svf,ATLAS:2022iyq,ALICE:2023plt}. The closest indication so far comes from a recent PHENIX measurement in d+Au collisions~\cite{PHENIX:2023dxl}.

Light-ion collisions, such as pO, OO, and Ne--Ne, offer a way to address this question. They reach multiplicities overlapping those of high-multiplicity pp collisions while retaining a nuclear geometry and a larger collision overlap region~\cite{Brewer:2021tyv,ALICE:OOprojections}. The first LHC light-ion run has already revealed features similar to those observed in heavy-ion collisions. These include measurements of anisotropic flow in OO and Ne--Ne collisions~\cite{ALICE:2025flow} and the suppression of neutral-pion production in OO collisions~\cite{ALICE:2026eloss}.

Di-hadron correlations are well suited for probing the system-size evolution of jet fragmentation~\cite{ALICE:2018hjy}. This method requires no jet reconstruction and remains applicable down to low $p_{\rm T}$, where medium-induced modifications are expected to be largest. The near-side peak at $\Delta\eta \approx 0$, $\Delta\varphi \approx 0$, where $\Delta\eta$ and $\Delta\varphi$ denote the relative pseudorapidity and relative azimuthal angle between the two particles of a pair, is primarily formed by pairs of particles originating from the fragmentation of the same hard-scattered parton. Interactions with the surrounding medium can alter the distribution of particles around the original parton direction, thereby modifying the width and shape of the peak. ALICE has reported an anomalous broadening of this peak along $\Delta\eta$ in Pb--Pb collisions at $\sqrt{s_{\rm NN}} = 2.76$~TeV, which grows continuously towards central collisions~\cite{ALICE:2016vzu,ALICE:2016ilo}, and a recent CMS measurement at $\sqrt{s_{\rm NN}} = 5.02$~TeV confirms a broadening that is much stronger in the longitudinal than in the azimuthal direction~\cite{CMS:2026jps}. Transport calculations attribute this asymmetry to the coupling between the propagating parton and the longitudinal flow of the medium~\cite{Ma:2008nd,Armesto:2004pt}. In pp collisions, only a mild narrowing towards high multiplicity is seen for high-$p_{\rm T}$ pairs~\cite{ALICE:2021jps,ALICE:2024jetmod}. This contribution reports the first such measurement in pO, OO, and Ne--Ne collisions.

\section{Analysis strategy}
\label{sec:anastat}
The OO and Ne--Ne collision samples were recorded during the 2025 light-ion run at $\sqrt{s_{\rm NN}} = 5.36$~TeV, the pO sample at $\sqrt{s_{\rm NN}} = 9.62$~TeV, and the pp reference sample at $\sqrt{s} = 5.36$~TeV in 2024. Events are selected with a minimum-bias trigger based on a coincidence between the two arrays, FT0A ($3.5 < \eta < 4.9$) and FT0C ($-3.3 < \eta < -2.1$), of the time-zero detector (FT0) of the Fast Interaction Trigger system~\cite{ALICE:2023upgrade}, with the primary vertex within $\pm 10$~cm of the nominal interaction point. The event activity is given by the FT0 amplitude, measured in FT0C for the nuclear systems and in both arrays for pp collisions, and the event-activity percentile classes are mapped onto the average charged-particle multiplicity at midrapidity, $\langle N_{\rm ch}\rangle$ ($|\eta| < 0.5$). Tracks used to construct the two-particle correlations are reconstructed in $|\eta| < 0.8$ and $1 < p_{\rm T} < 8$~GeV/$c$ with the Inner Tracking System and the GEM-based Time Projection Chamber~\cite{ALICE:ITSTDR,ALICE:TPCGEM}, following Ref.~\cite{ALICE:2025flow}.

Two-particle correlations are measured as a function of the relative pseudorapidity $\Delta\eta$ and azimuthal angle $\Delta\varphi$ of trigger and associated particles, requiring $p_{\rm T,\,trig} > p_{\rm T,\,assoc}$. The per-trigger normalised associated yield is calculated as
\begin{equation}
\frac{1}{N_{\rm trig}}\frac{{\rm d}^{2}N_{\rm pair}}{{\rm d}\Delta\eta\,{\rm d}\Delta\varphi} = B(0,0)\,\frac{S(\Delta\eta,\Delta\varphi)}{B(\Delta\eta,\Delta\varphi)},
\label{eq:corr}
\end{equation}
where $S$ is calculated from same-event pairs and $B$ from pairs combining a trigger particle with associated particles from five different events with similar multiplicity and vertex position. The mixed-event term corrects the shape of $S$ for the pair acceptance. Pair-by-pair efficiency corrections are applied, obtained from PYTHIA~8 (Monash tune~\cite{Skands:2014pea}) and Angantyr~\cite{Bierlich:2018xfw} simulations with GEANT3~\cite{Brun:1994aa}.

The peak width along $\Delta\eta$ ($\sigma_{\Delta\eta}$) is quantified by projecting the correlation function onto $\Delta\eta$ within $|\Delta\varphi| < 1.3$~rad and fitting it in $|\Delta\eta| < 1.3$ with a generalised Gaussian,
\begin{equation}
G_{\gamma,w}(x) = \frac{\gamma}{2w\,\Gamma(1/\gamma)}\exp\left[-\left(\frac{|x|}{w}\right)^{\gamma}\right] + A ,
\label{eq:ggaus}
\end{equation}
where $w$ and $\gamma$ are the scale and shape parameters, respectively, and $A$ accounts for the underlying event. The width of the distribution is given by $\sigma = w\,[\Gamma(3/\gamma)/\Gamma(1/\gamma)]^{1/2}$, with $\gamma=2$ corresponding to the standard Gaussian. The peak width along $\Delta\varphi$ ($\sigma_{\Delta\varphi}$) is obtained similarly by projecting onto $\Delta\varphi$ within $|\Delta\eta| < 1.3$. The pedestal and flow contributions are first subtracted, with the flow contribution estimated from the long-range $\Delta\varphi$ projections, and the resulting distribution is fitted with Eq.~\eqref{eq:ggaus}, in which $A$ is kept as a free parameter.
Systematic uncertainties are estimated in each $\langle N_{\rm ch}\rangle$ interval using the Barlow criterion~\cite{Barlow:2002yb}. The contributions come from variations of the vertex range and the event-activity estimator, the track selection, the fit range, and the simulation non-closure. In the figures, the systematic uncertainties are shown as boxes and the statistical uncertainties as vertical bars.

\section{Results}

\begin{figure}[!t]
    \centering
    \includegraphics[trim=0cm 0cm 0cm 0.2cm, clip,width=0.72\linewidth]{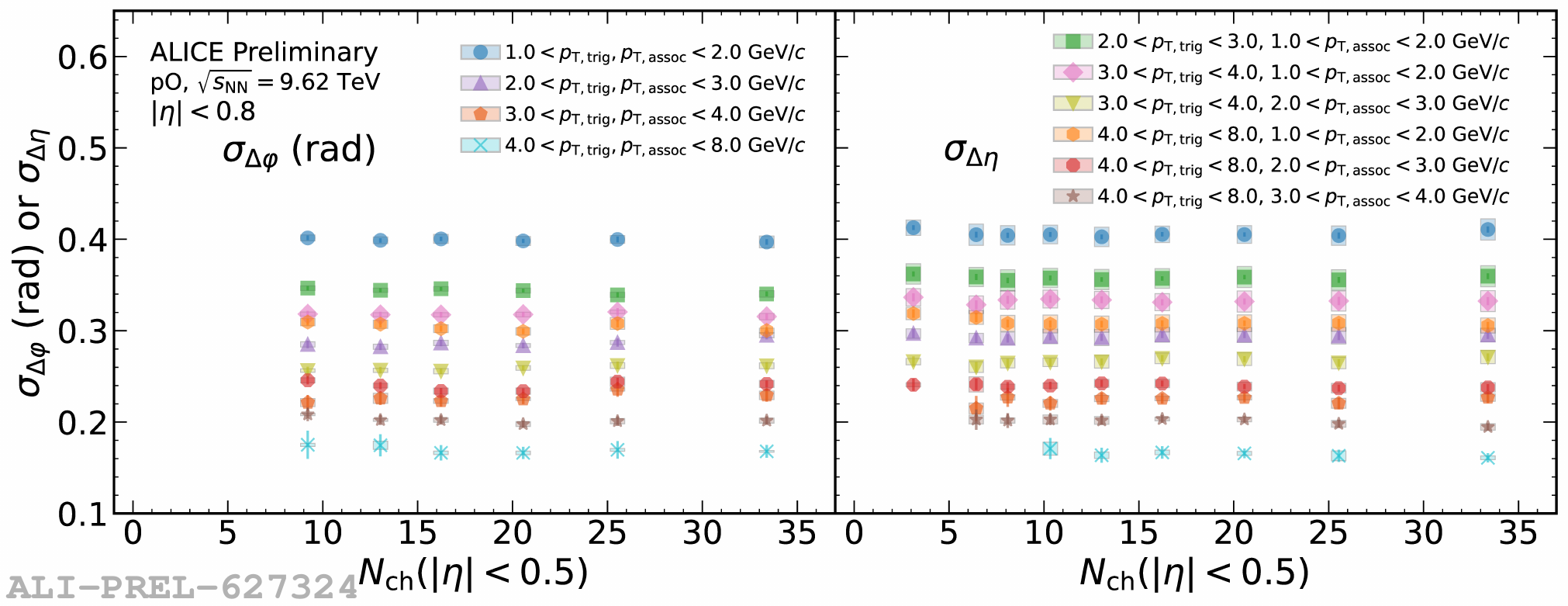}
    \caption{The azimuthal (left) and longitudinal (right) widths of the near-side peak in pO collisions at $\sqrt{s_{\rm NN}} = 9.62$~TeV as a function of $\langle N_{\rm ch}\rangle$ ($|\eta| < 0.5$), for different $p_{\rm T,\,trig}$ and $p_{\rm T,\,assoc}$ intervals.
    }
    \label{fig:pO}
\end{figure}

The results in pO collisions are presented first, as pO is the smallest of the light-ion systems and provides an intermediate reference between pp and the symmetric nuclear systems. Figure~\ref{fig:pO} shows the azimuthal and longitudinal near-side widths in pO collisions as a function of $\langle N_{\rm ch}\rangle$ ($|\eta| < 0.5$). The widths of the near-side peak along the azimuthal ($\sigma_{\Delta\varphi}$) and longitudinal ($\sigma_{\Delta\eta}$) directions are comparable. They decrease with increasing $p_{\rm T}$, as expected from the harder fragmentation of more energetic partons, but show no significant multiplicity dependence over the accessible range, $\langle N_{\rm ch}\rangle \lesssim 35$. The jet-like peak behaves as in pp collisions, with its width determined by fragmentation kinematics alone~\cite{ALICE:2024jetmod}.

\begin{figure}[!t]
    \centering
    \includegraphics[trim=0cm 0cm 0cm 0.1cm, clip, width=0.7\linewidth]{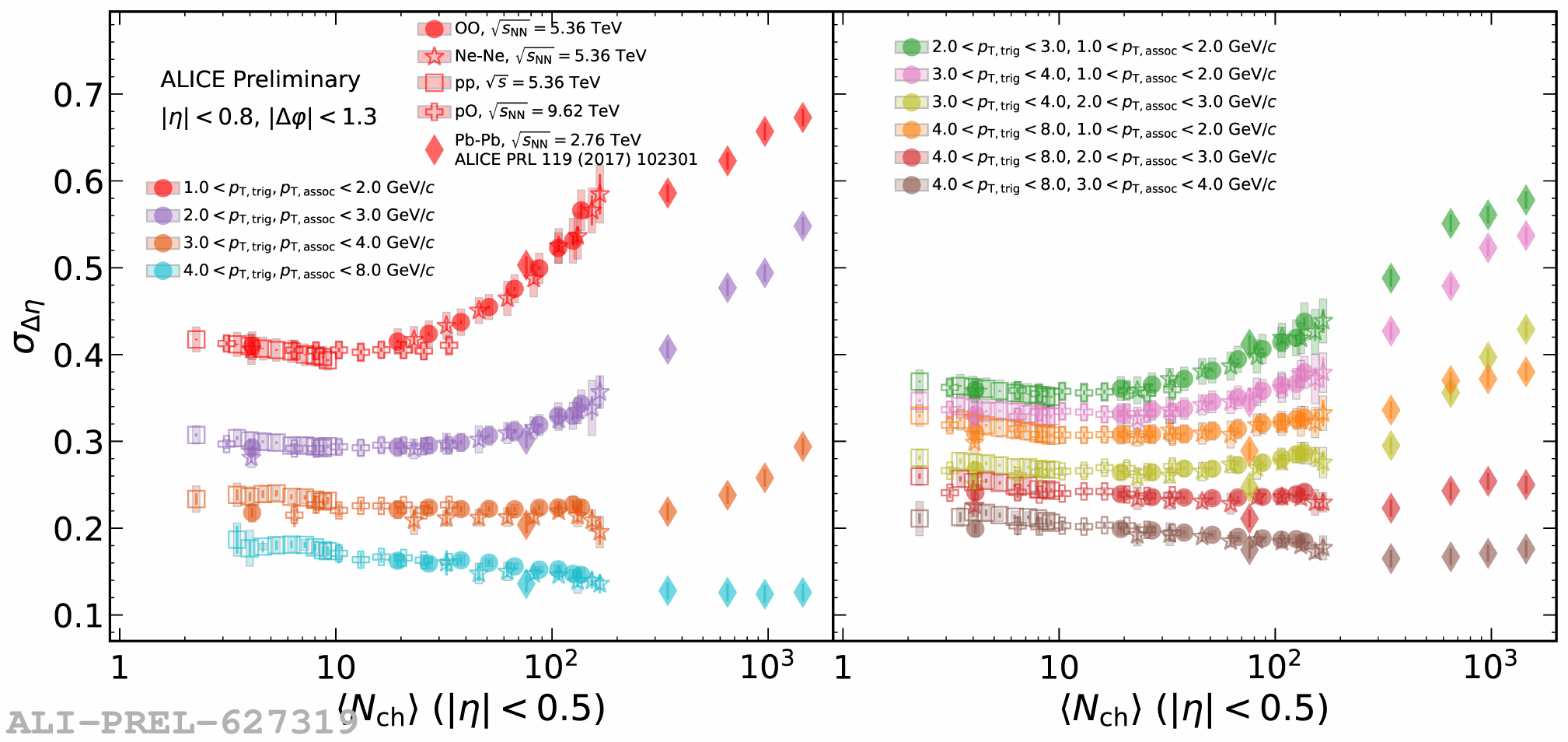}
    \caption{The longitudinal width of the near-side peak as a function of $\langle N_{\rm ch}\rangle$ ($|\eta| < 0.5$) in pp ($\sqrt{s} = 5.36$~TeV), pO ($\sqrt{s_{\rm NN}} = 9.62$~TeV), OO and Ne--Ne ($\sqrt{s_{\rm NN}} = 5.36$~TeV) collisions, compared with Pb--Pb collisions at $\sqrt{s_{\rm NN}} = 2.76$~TeV~\cite{ALICE:2016vzu}.}
    \label{fig:allsystems}
\end{figure}

The system-size dependence is shown in Fig.~\ref{fig:allsystems}, where $\sigma_{\Delta\eta}$ is presented as a function of $\langle N_{\rm ch}\rangle$ for all measured systems together with published Pb--Pb results~\cite{ALICE:2016vzu}. For $1 < p_{\rm T,\,trig},\,p_{\rm T,\,assoc} < 2$~GeV/$c$, the width is flat in pp and pO collisions but rises continuously in OO and Ne--Ne collisions above $\langle N_{\rm ch}\rangle \approx 20$. In central collisions, the width reaches values close to those measured in peripheral Pb--Pb collisions, with OO and Ne--Ne agreeing within uncertainties. The light-ion points connect smoothly to the Pb--Pb trend, so that all systems fall on one continuous curve spanning three orders of magnitude in $\langle N_{\rm ch}\rangle$. Multiplicity, rather than the colliding species, appears to be the relevant parameter controlling the near-side peak width.
The behaviour changes qualitatively for harder pairs. For $p_{\rm T,\,trig},\,p_{\rm T,\,assoc} \gtrsim 3$~GeV/$c$, the broadening disappears, and a slight narrowing with multiplicity is seen in all systems, including pp, although the trend is compatible with being flat within uncertainties. A similar narrowing was reported in pp collisions at $\sqrt{s} = 13$~TeV and interpreted as a bias towards harder fragmentation in high-multiplicity events~\cite{ALICE:2021jps,ALICE:2024jetmod}. The persistence of this narrowing in OO and Ne--Ne collisions suggests that it is largely an event-selection effect. In contrast, the broadening observed at low $p_{\rm T}$ appears to be specific to nuclear systems.

\begin{figure}[!t]
    \centering
    \includegraphics[trim=0cm 0cm 0cm 1.6cm, clip, width=0.75\linewidth]{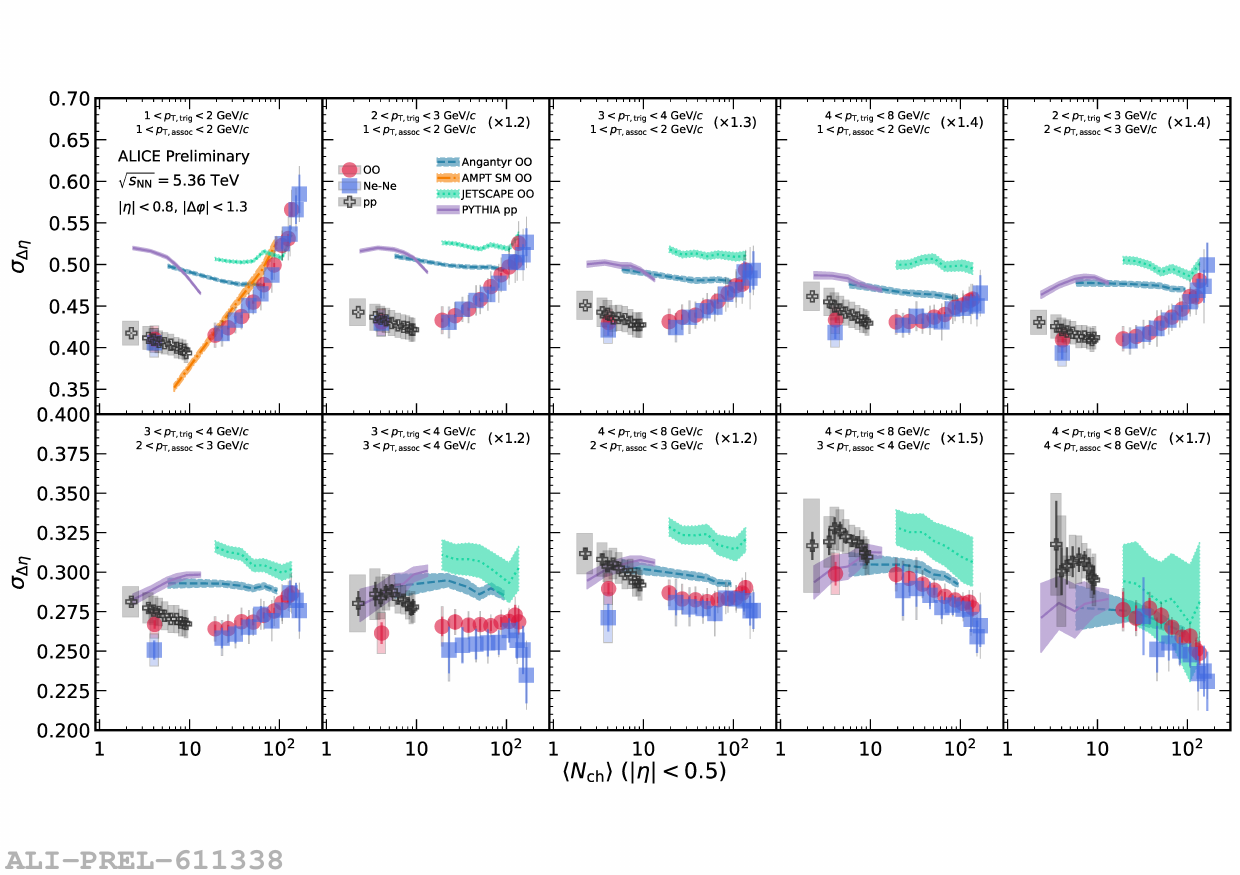}
    \caption{The multiplicity dependence of $\sigma_{\Delta\eta}$ in pp, OO, and Ne--Ne collisions for all measured $p_{\rm T,\,trig}$ and $p_{\rm T,\,assoc}$ combinations, compared with PYTHIA~8 (Monash tune for pp and Angantyr for OO), AMPT in string-melting mode, and JETSCAPE (OO only). The bands represent the statistical uncertainties of the model calculations. As indicated, some of the panels use a scaled vertical axis.}
    \label{fig:models}
\end{figure}

The data are compared in Fig.~\ref{fig:models} with three models differing in their treatment of jet--medium interactions. PYTHIA~8, with the Monash tune for pp and with Angantyr for nuclear collisions~\cite{Skands:2014pea,Bierlich:2018xfw}, serves as a vacuum reference without a hydrodynamic medium. AMPT in string-melting mode~\cite{Lin:2004en}, with a 3~mb parton cross section, adds a partonic stage, quark coalescence, and hadronic rescattering. JETSCAPE~\cite{Putschke:2019yrg} couples a PYTHIA~8 hard process to T\raisebox{-0.25ex}{\scriptsize R}ENTo initial conditions~\cite{Soeder:2023vdn}, MUSIC hydrodynamics~\cite{Schenke:2010nt}, and a multistage MATTER+LBT energy loss~\cite{Majumder:2013re,He:2015pra} with the jet transport coefficient $\hat{q}$ taken from Bayesian analyses of hadron suppression~\cite{JETSCAPE:2021ehl}. None of these Monte Carlo models describes the data over the full kinematic range. PYTHIA~8 with Angantyr, containing no medium evolution, predicts essentially no multiplicity dependence and fails at low $p_{\rm T}$, approaching the data only for $p_{\rm T,\,trig} > 4$ and $p_{\rm T,\,assoc} > 3$~GeV/$c$, where the measured dependence is weakest. AMPT reproduces both the magnitude and the rise of the width in the softest $p_{\rm T}$ regions, consistent with a final-state origin of the broadening, but overestimates the data for harder $p_{\rm T}$ pairs. JETSCAPE overestimates the width at low $p_{\rm T}$ and shows no multiplicity dependence, while reproducing the mild narrowing at high $p_{\rm T}$. Below 3~GeV/$c$, the width is sensitive to medium response and hadronisation, whereas JETSCAPE was tuned to high-$p_{\rm T}$ observables.

\section{Summary}
Jet-like di-hadron correlations are measured in pO, OO, and Ne--Ne collisions with ALICE. A significant broadening of the near-side peak towards high multiplicity is observed for soft pairs in OO and Ne--Ne collisions, which is absent in pp and pO collisions. At higher $p_{\rm T}$, the trend reverses into a mild narrowing common to all collision systems. The widths in pp, pO, OO, and Ne--Ne collisions are consistent with Pb--Pb results within uncertainties at similar $\langle N_{\rm ch}\rangle$, indicating that the near-side peak width evolves continuously with system size. Neither the vacuum reference nor the medium-based models reproduce the results across the full kinematic range.

\section*{Acknowledgements}
This work was supported by the Hungarian National Research, Development and Innovation Office (NKFIH) under Contract No. NKFIH ADVANCED\_25 K153456, NKFIH NEMZ\_KI-2022-00058, 2024-1.2.5-TET-2024-00022, and by the Wigner Scientific Computing Laboratory (WSCLAB, the former Wigner GPU Laboratory).

\end{document}